\documentstyle[12pt]{article}

\begin{document}

\title{A Predictive SO(10) Scheme for Fermion Masses and Mixings\footnote{Work
supported in part by Department of Energy Grant \#DE-FG02-91ER406267}}

\author{{\bf K.S.Babu and Q. Shafi}\\
{\it Bartol Research Institute}\\ {\it University of Delaware}\\
{\it Newark, DE 19716}}

\date{BA-95-06,hep-ph/9503313}
\maketitle

\begin{abstract}

\baselineskip=.23in

We present a predictive scheme for fermion masses and mixings inspired
by supersymmetric SO(10) in which the gauge hierarchy problem is
resolved without fine tuning the parameters.  There are six predictions
in the flavor sector, all consistent with the present data.  The scheme
reproduces the familiar asymptotic relations $m_b=m_\tau$ and
$m_dm_sm_b=m_e m_\mu m_\tau$.  A new expression for $V_{cb}$ is
obtained in terms of the quark masses.  The remaining predictions
involve the quark mixing angles $V_{us}$ and $V_{ub}$, as well as the
parameter tan$\beta$ which turns out to be close to $m_t/m_b$.

\end{abstract}
\newpage

\baselineskip=.23in

\section*{1. Introduction}

In this paper we present a scheme for fermion masses and mixings
motivated by supersymmetric $SO(10)$ grand unification (GUT)
in which the gauge hierarchy problem is overcome without fine tuning.
There are six
predictions in the flavor sector, all of which are
phenomenologically consistent and compatible with the existence of a
heavy top quark.  A simple derivation in $SO(10)$ of the quark and lepton mass
matrices that lead to these predictions will be presented.

The familiar asymptotic relations $m_b^0=m_\tau^0$ and
$m_d^0m_s^0m_b^0 = m_e^0m_\mu^0m_\tau^0$ (see [1])\footnote{We use the
superscript $^0$ to denote GUT scale quantities.}
are reproduced in our scheme, while a
new sum rule for $m_s^0$ in terms of the quark mixing
angle $V_{cb}$ is obtained.  Two of the remaining three predictions are
for $V_{us}$ and $V_{ub}$, resembling those from the Fritzsch
scheme [2], while the third prediction is for the ratio of the two
Higgs vacuum expectation values, tan$\beta \simeq m_t/m_b$ [3].

The model presented here provides a unified framework for
some of the well--known and successful relations for the quark
masses and mixing angles which are compatible with
a heavy top quark.  Since the up--quark,
down--quark and the charged--lepton mass matrices are assumed to
have similar forms, their derivation
from an underlying $SO(10)$ GUT is relatively straightforward.

The $SO(10)$ model is the simplest grand unified example
that assembles all fermions
of a given family into a single irreducible representation,
the spinorial {\bf 16}.
As such, it has proven useful to work with $SO(10)$
in attempting to model fermion masses and mixings.  It also has the
potential to
relate the neutrino masses and mixings to those of the quarks and
charged leptons [4].  The doublet--triplet mass--splitting
problem that all GUT models must address has an elegant
resolution in SUSY $SO(10)$ without any fine tuning of the
parameters [5,6,7].  We are also encouraged by recent attempts to derive
realistic supersymmetric $SO(10)$ GUTs from superstrings in the free
fermionic formulation [8,9].

In the $SO(10)$ model of fermion masses that we consider,
we shall require that the doublet--triplet
mass splitting problem be
resolved without fine tuning.  Furthermore,
motivated by our desire
to preserve the successful prediction of sin$^2\theta_W$,
we assume that $SO(10)$ breaks directly to the minimal supersymmetric
standard model (MSSM).  These two requirements typically call for a
non--minimal Higgs system at the GUT scale which we exploit to arrive at
a predictive scheme for quark and lepton masses and mixings.

\section*{2. Model}

We shall adopt the mechanism developed in Ref. [6,7] to solve the
doublet--triplet splitting problem naturally in SUSY $SO(10)$, making
use of an old suggestion of Dimopoulos and Wilczek [5].  This involves
the coupling of Higgs {\bf 10}'s with an adjoint {\bf 45}:
${\bf 10}_1. {\bf 45}. {\bf 10}_2$.  If the {\bf 45} acquires a
vacuum expectation value (VEV)
along the $(B-L)$ direction, $\left \langle {\bf 45} \right \rangle =
diag. (a,a,a,0,0) \times i\tau_2$, this coupling gives GUT scale masses
to the
color--triplets in the {\bf 10}'s, while leaving massless the
$SU(2)_L$ doublets to be identified with $H_u$ and $H_d$ of
MSSM.  The superpotential term $(M_{\rm GUT} {\bf 10}_2^2)$
will make the doublets in
${\bf 10}_2$ superheavy, leaving MSSM as the effective low energy theory.

The minimal Higgs sector that can break the $SO(10)$ gauge symmetry
down to MSSM
involves one {\bf 45} and a spinorial ${\bf 16} +
\overline{\bf 16}$ [7].  However, the low energy
theory below the GUT scale in this case is not quite MSSM.  In
particular, some of the
Higgs(ino) superfields
turn out to have intermediate scale masses.
Here, since we require that the
theory below the GUT scale be MSSM,
it will be necessary to have a slightly more extended
Higgs system at the GUT scale.

The Higgs system that we employ for the symmetry breaking involves two
sectors.  One sector breaks
$SO(10)$ down to $SU(5)$.  This can be achieved for example, via the
following superpotential involving a {\bf 45} ($A$) and a ${\bf 16}+
\overline{\bf 16}$ ($C+\overline{C}$):
\begin{equation}
W_1 = m_1 \overline{C} C + m_2 A^2 + \lambda_1 \overline{C}CA~.
\end{equation}
This superpotential induces a VEV for $C$ and $\overline{C}$ along
the $SU(5)$ singlet direction, and for $A$ along $SU(5) \times U(1)_X$:
$\left \langle A \right \rangle = diag.(a,a,a,a,a,) \times
i\tau_2$.

The second sector breaks $SO(10)$ down to $SU(3)_c \times SU(2)_L
\times SU(2)_R \times U(1)_{B-L}$.  The superpotential involves
a {\bf 54} ($S$) and a second adjoint $A^\prime$ [10]:
\begin{equation}
W_2 = m_3 S^2 + m_4 A^{\prime 2} + \lambda_2 S^3 + \lambda_3 A^{\prime 2
}S~.
\end{equation}
The resulting VEVs of $S$ and $A^\prime$ are $\left \langle S
\right \rangle = diag.(s,s,s,-{3 \over 2} s, -{3 \over 2}s) \times 1$,
$\left \langle A^\prime \right \rangle = diag. (a',a',a',0,0)\times
i\tau_2$.  The $A^\prime$ field in this sector is also responsible for
the doublet--triplet splitting via the superpotential terms involving
Higgs {\bf 10}--plets (denoted by $H$ and $H^\prime$)
\begin{equation}
W_3 = H A^\prime H^\prime + m_5 H^{\prime 2}~.
\end{equation}

In order to keep the masses of the electroweak Higgs doublets at the
correct scale, the
VEV of $A^\prime$ should be along $(B-L)$ to a high degree of
accuracy.  This requires that the $A^\prime$ coupling to the $(A, C)$ sector
be very weak, since any such coupling will induce an $A^\prime$ VEV
along the $SU(5) \times U(1)_X$ singlet direction as well.  Now
the $(S, A^\prime)$ sector should be linked with the $(A,C)$ sector,
otherwise pseudo-Goldstone bosons will result.  The simplest way to
link the two sectors without upsetting the VEV pattern of $A^\prime$
is [6] by a coupling involving another adjoint:
Tr$(A A^\prime A^{\prime \prime})$.
This term, due to its complete antisymmetry, vanishes at the minimum,
and thus does not affect the VEV of $A^\prime$.  Yet, it gives GUT
scale masses to all the would-be pseudo-Goldstone bosons.  The
$A^{\prime \prime}$ superfield can have its own superpotential
and its VEV can in general
be written as $\left \langle A^{\prime \prime} \right \rangle =
diag.a^{\prime \prime}(1+z, 1+z, 1+z,
1-{3\over 2} z, 1-{3\over 2}z)\times i\tau_2$.  Here $z=0$ would
correspond to the VEV being along $U(1)_X$, while $z=-1~(2/3)$ would preserve
$I_{3R}~ (B-L)$.

Since our focus
here is the fermion Yukawa sector, we shall not get into
the details of symmetry breaking, except to note that one can find
discrete symmetries that would lead to the
superpotential given in Eqs. (1)-(3) along with the Tr$(AA^\prime
A^{\prime \prime})$ term, while preventing other terms which potentially
can upset the gauge hierarchy.  Note that one has
the option of using gauge singlet superfields with GUT scale
VEVs to induce some of the mass terms in Eqs. (1)-(3).

We now proceed to show that the above Higgs structure can lead to a
predictive set of
mass matrices for the quarks and charged leptons, provided one introduces a
flavor--dependent
discrete symmetry.

\section*{3. Fermion Mass Matrices}

In our scheme, only the third family receives mass
from a renormalizable operator in the superpotential.  The
second family masses as well as the mixing angle arise
from dimension 4 operators in the
superpotential, suppressed by one inverse power of $M$, where
$M$ is a scale much larger than the GUT scale.  The first family masses
and mixings will come from dimension 5 operators.  It is entirely
conceivable that these non--renormalizable
operators arise from integrating out some vector--like families
with masses of order $M$.
The inter--generational mass hierarchy is thus related to the small
ratio $(M_{\rm GUT}/M)$, an idea discussed by numerous authors in the
past [11].

In our discussions we shall not assume any special
structure for these `effective' operators.
They will only be constrained by the $SO(10)$ gauge symmetry and some
discrete flavor symmetry.  In
particular, if one contraction of the group indices is allowed in a given
nonrenormalizable operator, we shall allow for
all possible group contractions.
This should make the derivation of such `effective' operators from an
underlying theory somewhat easier, since no special care is needed in the
way the heavy fields are integrated out.

As indicated above, the third family acquires its mass from a
direct coupling to the
Higgs {\bf 10}--plet $H$.  All the other masses will involve
the adjoint fields $A$ and $A^{\prime \prime}$ with VEVs specified
earlier.  In order to generate the first family masses and mixings, a
singlet field $X$ with a GUT scale VEV
is also employed.  The Yukawa superpotential is given by
\begin{eqnarray}
W_{\rm Yuk} &=& h_{33} \psi_3^T C \gamma_a \psi_3 H_a +
h_{23} \psi_2^T C \gamma_{\{abc\}}\psi_3A^{\prime \prime}_{ab}H_c
+ h_{23}^\prime \psi_2^T C \gamma_a \psi_3 A^{\prime \prime}_{ab} H_b
\nonumber \\
&+& h_{12}\psi_1^T C \gamma_{\{abc\}}\psi_2A_{ab}H_cX +
h_{12}^\prime \psi_1^T C \gamma_a \psi_2 A_{ab}H_bX~.
\end{eqnarray}
Here $\psi_i$, $(i=1-3)$ stand for the three fermion families
belonging to the {\bf 16} of $SO(10)$, $C$ is the $SO(10)$ charge
conjugation matrix, and $\gamma_{\{abc\}}$ is the totally antisymmetric
combination of the $SO(10)$ gamma matrices:
\begin{equation}
\gamma_{\{abc\}} = \gamma_a \gamma_b \gamma_c - \gamma_a \gamma_c
\gamma_b - \gamma_b \gamma_a \gamma_c + \gamma_c\gamma_a\gamma_b +
\gamma_b\gamma_c\gamma_a-\gamma_c\gamma_b\gamma_a ~.
\end{equation}
In Eq. (4), appropriate powers of $M$ in the denominators are to
be understood.

As an example of a discrete symmetry that would lead to the Yukawa terms
of Eq. (4), consider the transformations of the relevant fields
under $z_5$:
$(\psi_1,\psi_2,\psi_3) \sim
(\omega^2,\omega,1)$, $A^{\prime \prime} \sim
\omega^4, X \sim \omega^2$, where $\omega^5=1$.  This guarantees the
radiative stability of the Yukawa couplings.
Note that in the higher dimensional operators in Eq. (4),
both the {\bf 10} and {\bf 120} `effective' operators contribute, i.e.,
all possible contraction of the $SO(10)$ group indices have been included.

The couplings $h_{33}, h_{23}, h_{12}$ can be made real by field
redefinitions, while $h_{12}^\prime, h_{23}^\prime$ and $z$ are in general
complex parameters.

To arrive at the charged fermion mass matrices from Eq. (4),
it is necessary to determine
the couplings of all the components of ${\bf 16}_i$ with the light doublets
($H_u,H_d$) in $H$.  This is carried out by
adopting a definite set of $SO(10)$ gamma matrices [12].  The fields
$H_u$ and $H_d$ in the notation of Ref. [12] are $H_u=(H_9+iH_{10})/\sqrt{2}$
and $H_d = (H_9-iH_{10})/\sqrt{2}$.  Consequently, the $h_{23}$ term in
Eq. (4) can be expanded as (after absorbing $a^{\prime \prime}/M$ into
$h_{23}$, and similarly $\left \langle AX \right \rangle/M^2$ into
$h_{12}$)
\begin{eqnarray}
&~& \sqrt{2}ih_{23}H_u \left[u_2u_3^c(-Q_u+2Q_{u^c}-(1-{3\over 2}z))
+u_3u_2^c(-Q_{u^c}+2Q_u-(1-{3\over 2}z))\right] \nonumber \\
&+& \sqrt{2}ih_{23}H_u\left[\nu_2\nu_3^c(-Q_\nu+2Q_{\nu^c}-(1-{3 \over 2
}z)) + \nu_3\nu_2^c(-Q_{\nu^c}+2Q_\nu-(1-{3 \over 2}z))\right] \nonumber
\\
&+&  \sqrt{2}i h_{23}H_d \left[d_2d_3^c(-Q_d+2Q_{d^c}+(1-{3\over 2}z))+
d_3d^c_2(-Q_{d^c}+2Q_d+(1-{3\over 2}z))\right]\nonumber \\
&+& \sqrt{2}ih_{23}H_d \left[e_2e_3^c(-Q_e+2Q_{e^c}+(1-{3\over 2}z))+
e_3e_2^c(-Q_{e^c}+2Q_e+(1-{3\over 2}z))\right] \nonumber \\
\end{eqnarray}

\noindent Similarly, the $h_{23}^\prime$ term gives rise to
\begin{equation}
\sqrt{2}ih_{23}^\prime (1-{3\over 2}z)\left[-H_u(u_2u_3^c+u_3u_2^c+\nu_2
\nu_3^c+\nu_3\nu^c_2)+
H_d(d_2d_3^c+d_3d_2^c+e_2e_3^c+e_3e_2^c)\right]~.
\end{equation}
Here we have defined the ``charges'' of quarks and leptons to be
$Q = X + 6 \left({Y \over 2}\right)$, where $X$ is the $U(1)_X$ charge
normalized such that the {\bf 16} of $SO(10)$ decomposes into $SU(5) \times
U(1)_X$ as ${\bf 16} \rightarrow {\bf 1}_{5} + {\bf \overline{5}}_{-3}+
{\bf 10}_1$.
These charges are given by
\begin{eqnarray}
Q_u &= & 1+z, ~Q_{u^c} = 1-4z,~ Q_d=1+z,~ Q_{d^c} = -3+2z, \nonumber \\
Q_e &= & -3-3z,~Q_{e^c}=1+6z, ~Q_\nu = -3-3z,~ Q_{\nu^c} = 5~.
\end{eqnarray}

We can now write down the mass matrices for the up--quark, down quark
and charged leptons.  Let us define $a_{23} \equiv \sqrt{2}ih_{23}$,
$r_{23}a_{23} \equiv \sqrt{2}ih_{23}^\prime(1-{3\over 2}z)$, $a_{12}
\equiv
\sqrt{2}ih_{12}, r_{12} \equiv \sqrt{2}ih_{12}^\prime$.  The mass
matrices are then given by
\begin{eqnarray}
M_{u} &=& v_u \left(\matrix{0 & -a_{12}r_{12} & 0 \cr
-a_{12}r_{12} & 0 & -a_{23}({15 \over 2}z+r_{23}) \cr
0 & -a_{23} (-{15 \over 2}z+r_{23}) & a_{33}}\right) \nonumber
\end{eqnarray}
\begin{eqnarray}
M_d &=& v_d \left(\matrix{0 & a_{12}(-6+r_{12}) & 0 \cr
a_{12}(6+r_{12}) & 0 & a_{23}(-6+{3 \over 2}z+r_{23}) \cr
0 & a_{23}(6-{3 \over 2}z+r_{23}) & a_{33}}\right) \nonumber
\end{eqnarray}
\begin{eqnarray}
M_l &=& v_d \left(\matrix{0 & a_{12}(6+r_{12}) & 0 \cr
a_{12}(-6+r_{12}) & 0 & a_{23}(6+{27 \over 2}z+r_{23}) \cr
0 & a_{23}(-6-{27\over 2}z+r_{23}) & a_{33}}\right)
\end{eqnarray}

A few remarks about (9) are in order.  By
construction, they have a Fritzsch--like texture.  However, there is an
important difference in that the matrices are not symmetric.  This is because
both the symmetric {\bf 10} and the antisymmetric {\bf 120}
`effective' operators contribute to the
(12) and (23) entries in (9).  The net
contribution is then neither symmetric nor antisymmetric.
This implies in particular that the top--quark mass in our present
scheme can be larger than the 150 GeV limit [13] set by the Fritzsch
ansatz (the limit is $\le 90~GeV$ without the renormalization group
considerations [14]).
The matrices in (9) admit a top quark as heavy as about 200 GeV.

Another point to note is that the VEV of $A^{\prime \prime}$, being
proportional to the parameter $z$, breaks the $SU(5)$ symmetry.  Thus the bad
predictions of minimal $SU(5)$, namely $m_s^0=m_\mu^0$ and $m_d^0=m_e^0$
will be corrected.

As noted earlier, the parameters ($z,r_{12},r_{23}$) are in
general complex.  In the analysis that follows, we shall assume that
CP is a good symmetry of the Lagrangian.  It can be
spontaneously broken
by the VEV's of $A,A^{\prime \prime}$ or $X$.  It is easy
to see that if CP is a good symmetry of the superpotential in (1)-(3),
it admits a solution where $\left \langle A \right \rangle$ and
$\left \langle A^{\prime \prime} \right \rangle$ are real.
If this solution is chosen, the only
possible source of CP violation in the mass matrix
is in the phase of the singlet
field $X$.  If two such fields are present, there is,
in general, a phase in $r_{12}$ which cannot be rotated away.
We shall assume in the analysis that follows that $r_{12}$ is complex,
while $(z,r_{23})$ are real.

The assumption of spontaneous CP violation is of course motivated by
the desire to reduce the number of arbitrary parameters.  Within the
context of supersymmetry, there is another reason.  It is a well known
problem that the new phases in the soft SUSY breaking sector of the
MSSM, unless somehow suppressed, will lead to unacceptably
large values of the neutron and electron electric dipole moments.  One
way to suppress these potentially dangerous contributions is to assume
that CP violation has a spontaneous origin [15].

With only $r_{12}$ complex, there are 8 parameters ($a_{33}, a_{23},
a_{12}, r_{23}, r_{12}, \alpha, z, v_u/v_d)$ in Eq. (9) ($\alpha$ is
the phase of $r_{12}$) to fit 14 observables (6 quark masses, 3 lepton
masses, 3 quark mixing angles, one CP phase and tan$\beta$).  This
results in 6 predictions in the flavor sector as advocated.

Two of the predictions of the model are
\begin{equation}
m_b^0 \simeq m_\tau^0;~~~{\rm tan}\beta \simeq {{m_t}\over {m_b}}
\end{equation}
which can be seen by considering the (33) elements of $M_{u,d,l}$.
The resulting $b$--quark mass at low energies is known to be consistent
with data [3].

Another prediction of the model is
\begin{equation}
m_d^0m_s^0m_b^0 = m_e^0m_\mu^0m_\tau^0
\end{equation}
which follows from the determinants of $M_d$ and $M_l$.  This is one of
the Georgi--Jarlskog relations [1], which also fits the low energy data quite
well.

The remaining predictions have to do with the three quark mixing
angles.  The first one relates $V_{cb}$ to ratios of
quark and lepton masses.  This relation is new, and so we shall explain its
derivation in some detail.
To arrive at it one can safely ignore the masses and mixings involving
the first family.  Define
\begin{eqnarray}
P \equiv \eta_P \left|{{m_c^0m_b^0}\over {m_t^0m_s^0}}\right|;~~
Q \equiv \eta_Q \left|{{m_s^0m_\tau^0}\over {m_b^0m_\mu^0}}\right|
\end{eqnarray}
where $\eta_{P,Q} = \pm 1$, depending on the signs of the
fermion masses.  One can see that
\begin{eqnarray}
z&=&{{2(1-Q)}\over {3+7Q-10PQ}}~, \\
r_{23}^2 &=& {{225\left[(1+Q-4PQ)^2-4Q\right]}\over{(3+7Q-10PQ)^2}}~.
\end{eqnarray}
Numerically, the parameters $|P|$ and $|Q|$ lie in the range
\begin{equation}
|P| = \left({1 \over 5.4} ~{\rm to}~ {1 \over 7.0}\right);~|Q| =
\left({1 \over 2.9}~
{\rm to}~{1 \over 3.7}\right)
\end{equation}
where we have used the relation $m_s^0m_d^0 = m_\mu^0m_e^0$,
extrapolated the light quark masses from low energies to the top
mass scale ($\sim 160~ GeV-190~GeV)$ using $\alpha_s(M_Z) = 0.12$,
and imposed the phenomenological constraint $m_s/m_d= (15$ to $25$).

Since $r_{23}^2 > 0$,
\begin{equation}
|1+Q-4PQ| \ge 2\sqrt{|Q|}~.
\end{equation}
Depending on the sign factors
$\eta_P$ and $\eta_Q$ there are then four possibilities,

(i) For $\eta_p =+, \eta_Q=+$, Eq. (16) translates into an
inequality\footnote{For brevity, we denote $m_t^0$ by $t^0$ and so on.}
\begin{equation}
\left|{{t^0}\over {c^0}}\right| \ge  {{4|{{\tau^0}\over {\mu^0}}|}\over
{\left(1-\left|{{e^0s^0}\over{\mu^0d^0}}\right|^{1/4}\right)^2}}~.
\end{equation}
This leads to a lower limit on $m_t^{\rm phys}$ of about 165
GeV.

(ii) If $\eta_P = -, \eta_Q = -$, then
\begin{equation}
|P| \le {{(1-|Q|-2\sqrt{|Q|})}\over {4|Q|}}
\end{equation}
which cannot be satisfied since the right hand side is negative.

(iii) If $\eta_P = +, \eta_Q = -$,
\begin{equation}
\left|1-|Q|+4|PQ|\right| \ge 2\sqrt{|Q|}~,
\end{equation}
leading to the constraint
\begin{equation}
\left|{{t^0}\over {c^0}}\right|
\le {{4\left|{{\tau^0}\over {\mu^0}}\right|}
\over {\left(2\left|{{e^0s^0}\over {\mu^0d^0}}\right|^{1/4}+\left|{{e^0s^0}
\over{\mu^0d^0}}\right|^{1/2}-1\right)}}~.
\end{equation}
This results in an upper limit on $m_t^{\rm phys}$ of about 125 GeV
which is inconsistent with the data.

(iv) If $\eta_P=-, \eta_Q = +$, Eq. (16) leads to
the inequality $1+|Q|+4|PQ| \ge 2\sqrt{|Q|}$, which is
automatically satisfied.

The allowed solutions therefore are
$\eta_P = \pm$, $\eta_Q = +$.
Using these sign factors, the asymptotic expression for $V_{cb}$ can be
readily
obtained.  It is given by the relation $|V_{cb}^0| = |a_{23}/a_{33}| |6-
9z-2r_{23}|$, with $|a_{23}/a_{33}| = |c^0/t^0|^{1/2}/
|r_{23}^2-(15z/2)^2|^{1/2}$.  Using (13),(14), this yields ($x \equiv
|e^0s^0/\mu^0d^0|^{1/2}$, $y \equiv |c^0\tau^0/t^0\mu^0|$)
\begin{eqnarray}
\left|V_{cb}^0\right|=
{1 \over \sqrt{2}} \left| {\mu^0 \over \tau^0} \right|^{1/2}
{\left[2x-2\eta_P y \pm \left((1+x-4\eta_P y)^2-4x \right)^{1/2}
\right] \over (1+x-2\eta_P y)^{1/2} }~.
\end{eqnarray}
Here we have used the relation $|m_s^0/m_\mu^0|=
|{{m_e^0}\over {m_\mu^0}}{{m_s^0}\over
{m_d^0}}|^{1/2}$ which is a consequence of the identity $|m_d^0m_s^0| =
|m_e^0m_\mu^0|$.  The $\pm$ sign in (21) corresponds to choosing
$r_{23}$ to be $\mp$.  Only the positive sign for $r_{23}$
will lead to an acceptable $V_{cb}$.

It is instructive to approximate (21) in the limit of an infinite
top--quark, even though the finite top mass effects will turn out to be
significant.  In this limit, $|V_{cb}^0|$ can be written as (for $r_{23}
$ positive)
\begin{equation}
|V_{cb}^0| \simeq {1 \over \sqrt{2}} {\left|{{\mu^0}\over
{\tau^0}}\right|^{1/2}}
{1 \over {\sqrt{1+x}}} \left[(3x-1)+\eta_P \left|{{c^0 \tau^0} \over
{t^0\mu^0}}\right|\left({{1+12x+3 x^2}\over {1-x^2}}\right)\right]~.
\end{equation}
Note that in the strict
Georgi--Jarlskog limit, viz., $|\mu^0|=3|s^0|,~|d^0| = 3|e^0|$,
$x=1/3$ and the dominant $(3x-1)$ term in (22) vanishes.  As a
consequence, the model
admits very low values of $|V_{cb}|$.

Another useful expression for $|V_{cb}^0|$
which resembles the Fritzsch
relation is
\begin{equation}
|V_{cb}^0| = \left|\sqrt{ {s^0} \over {b^0} }\left|{6-{3 \over 2}z-
r_{23} \over 6 - {3 \over 2}z+r_{23} } \right|^{1/2} - \sqrt{ c^0
\over t^0 }\left|{ {15 \over 2}z+r_{23} \over {15 \over 2}z-r_{23}}
\right|^{1/2}\right|~.
\end{equation}
It is evident from (23) that for $r_{23} \ge 0$, the $\sqrt{s^0/b^0}$
term has a suppression factor while the $\sqrt{c^0/t^0}$
term is enhanced ($z \simeq +1/4$), thereby yielding
values of $|V_{cb}^0|$ that are significantly smaller than those from
the Fritzsch ansatz.

Relation (21) for $|V_{cb}^0|$ can be extrapolated to low energies by using the
renormalization group equations corresponding to tan$\beta \simeq
m_t/m_b$.  The running factors for the relevant quantities to go
from the weak to the GUT scale can be expressed analytically as
(tan$\beta\simeq m_t/m_b$)
\begin{eqnarray}
\eta_{KM} &=& \left(1-{{Y_t}\over {Y_f}}\right)^{1/7};~
\eta_{s/b}= \eta_{c/t} = \left(1-{{Y_t}\over {Y_f}}\right)^{2/7}; \nonumber \\
\eta_{\mu/\tau} &=& \left({{\alpha_1}\over {\alpha_G}}\right)^{1879/37224}
\left({{\alpha_3}\over {\alpha_G}}\right)^{2/3} \left({{Y_t}\over
{Y_\tau}}\right)^{-3/8} \left(1-{{Y_t}\over {Y_f}}\right)^{3/14}~.
\end{eqnarray}
Here $Y_t = h_t^2$ is the square of the top-quark Yukawa coupling, $Y_f$
($\simeq 1.1$) is the weak scale value of $Y_t$ corresponding to
`infinite' value at the
GUT scale, and $Y_\tau$ is the weak scale value
of the $\tau$ Yukawa coupling--squared.

The predicted low energy values for $|V_{cb}|$ are plotted in Figures 1 and 2
(corresponding to $\eta_P = \pm 1$)
as a function of the mass ratio $|m_s/m_d|$.
We have used the three-loop QCD and 1 loop QED beta and gamma
functions to evolve the light quark masses
from low energies to the top-quark threshold
(which is also assumed to be the SUSY threshold).  These running
factors, corresponding to $\alpha_s(M_Z) = 0.12$ are
$(\eta_u,\eta_{d,s},\eta_c,\eta_b,\eta_{e,\mu},\eta_\tau) =
(0.401,0.404,0.460,0.646,0.982,0.984)$, where the low energy scale is taken to
be 1 GeV for $(u,d,s)$, 1.27 GeV for $c$, and 4.25 GeV for $b$.
In the
square root in Eq. (21), only the relative negative sign (corresponding
to $r_{23} \ge 0$) leads to an
acceptable value of $|V_{cb}|$.  This relative minus sign has been used
in Figures 1 and 2.  Clearly, both signs of $\eta_P$ result in
a consistent prediction for $|V_{cb}|$.  A `central' value of
$V_{cb}=0.04$ corresponds to the prediction $m_s/m_d=(18,19.5,24)$
for $\eta_P =+1$ and $m_s/m_d=(22,21,17.5)$ for $\eta_P =-1$,
with $m_t^{\rm phys} = (165,174,190)$ GeV respectively.

Finally, the remaining two
mixing angle predictions of the model are
\begin{eqnarray}
|V_{us}^0| &\simeq & \left|\sqrt{{d^0 \over s^0}}\left(1-\left|{b^0\over
t^0}\right|
\left|{{c^0u^0}\over {d^0s^0}}\right|^{1/2}e^{i\alpha}\right) - \eta_P
e^{i\alpha} \sqrt{{u^0 \over c^0}}\right| ~,\nonumber \\
|V_{ub}^0| &=& \left|\sqrt{ {u^0\over c^0} }V_{cb}^0 - e^{i \alpha} {s^0 \over
b^0 }\sqrt{ {d^0 \over b^0} }\left|{ 6-r_{12} \over 6+r_{12}}
\right|^{1/2} \left|{6-{3 \over 2}z+r_{23} \over 6-{3 \over 2}z-r_{23}}
\right|^{1/2}\right|~.
\end{eqnarray}
In the expression for $|V_{ub}^0|$, the second term is numerically
smaller than the
first term, although not negligible.  Taking the phase
$\alpha \simeq \pi/2$, we find that $|V_{ub}| \simeq (0.002-0.003)$.
This and
the $|V_{us}|$ prediction resemble those from the
Fritzsch ansatz, and are quite consistent with the present data.  The CP
parameter turns out to be $J \approx (2-3)
\times 10^{-5}$.

\section*{4. Conclusions}

We have presented a scheme for fermion masses inspired by supersymmetric
$SO(10)$ in which the gauge hierarchy is implemented without
fine tuning.  This typically calls for a non--minimal Higgs sector
which we exploit in deriving expression (9) for the mass matrices.
Assuming spontaneous CP violation, we
are led to 6 predictions in the flavor sector which
work very well, especially with a heavy top quark.

There exist several ansatzes for the fermion mass matrices
in the literature [16-21].
The Fritzsch ansatz is one of the simplest and therefore attractive.
However, it would appear [13] that this ansatz is excluded by
the recent Fermilab data on the top quark mass.  The
Georgi--Jarlskog mass matrices have generated renewed
attention [16-17].  They lead to
6 predictions in the flavor sector.
Owing to the up--down asymmetry in this scheme, the derivation of these
mass matrices from an underlying theory
is somewhat nontrivial [17].  In comparison, the scheme
presented here can be obtained
from SUSY GUTs without too much effort.  There are some approaches
with more than six flavor predictions [19], and it would be interesting to see
the realization of these matrices from an underlying GUT or related
symmetries.

As for the neutral sector, small
neutrino masses can be easily accommodated in our scheme.  The
Dirac neutrino mass matrix is completely determined in the model
(see Eq. (6)).  As for the right handed ($\nu_R$)
Majorana mass matrix, we find that there are three
choices which lead to a predictive neutrino spectrum [4,21].
They correspond to
the non--zero entries in the Majorana matrix
being $\{(11), (23), (32)\}$, or $\{(22), (13),
(31)\}$ or $\{(33), (12), (21)\}$, all of which result in a non--singular
$\nu_R$ matrix.  In each case there is a one--parameter family of
solutions for the neutrino mass ratios and the mixing angles.  We plan
to discuss the detailed phenomenology of such spectra and their
implications for neutrino oscillations in a separate paper.

\section*{Acknowledgments}

We thank Steve Barr for discussions.

\section*{References}
\begin{enumerate}
\item H. Georgi and C. Jarlskog, Phys. Lett. {\bf 86B}, 297 (1979).
\item H. Fritzsch, Phys. Lett. {\bf 70B}, 436 (1977); Nucl. Phys.
{\bf B155}, 189 (1979); K.S. Babu and Q. Shafi, Phys. Rev.
{\bf D 47}, 5004 (1993).
\item B. Ananthanarayan, G. Lazarides and Q. Shafi, Phys. Rev. {\bf D44},
1613 (1991); M. Bando, T. Kugo, N. Maekawa and H.
Nakano, Mod. Phys. Lett. {\bf A 7}, 3379 (1992);
M. Carena, M. Olechowski, S. Pokorski and C. Wagner,
Nucl. Phys. {\bf B426}, 269 (1994); L.J. Hall, R. Rattazzi, and U.
Sarid, Phys. Rev. {\bf D 50}, 7048 (1994); V. Barger et. al., Phys. Rev.
{\bf D 47}, 1093 (1993); P. Langacker and N. Polonsky, $ibid$.,
{\bf 49}, 1544 (1994); B. Ananthanarayan, Q. Shafi and X.M. Wang,
$ibid$., {\bf 50}, 5980 (1994); M. Olechowski and S. Pokorski, Phys.
Lett. {\bf B214}, 393 (1988).
\item G. Lazarides and Q. Shafi, Nucl. Phys. {\bf B350}, 179 (1991);
K.S. Babu and R.N. Mohapatra, Phys. Rev. Lett. {\bf 70}, 2845
(1993); A.Yu. Smirnov, Phys. Rev. {\bf D 48}, 3264 (1993).
\item S. Dimopoulos and F. Wilczek, Report No. NSF-ITP-82-07, August
1981 (unpublished); R.N. Cahn, I. Hinchliffe and L. Hall, Phys. Lett.
{\bf 109B}, 426 (1982).
\item K.S. Babu and S.M. Barr, Phys. Rev. {\bf D 48}, 5354 (1993); Phys.
Rev. {\bf D 50}, 3529 (1994).
\item K.S. Babu and S.M. Barr, Bartol Preprint BA-94-45 (1994),
hep-ph/9409285 (to be published in Phys. Rev. D).
\item D. Lewellen, Nucl. Phys. {\bf B337}, 61 (1990); A. Font, L. Ibanez,
and F. Quevedo, Nucl. Phys. {\bf B345}, 389 (1990).
\item S. Choudhuri, S. Chung, G. Hockney, and J. Lykken, hep-ph/9501361,
Fermilab-Pub-94/413-T (1994); G. Cleaver, hep-th/9409096, Preprint
OHSTPY-HEP-T-94-007 (1994); G. Aldazabal, A. Font, L. Ibanez and A.
Uranga, hep-th/9410206, Preprint FTUAM-94-28 (1994).
\item M. Srednicki, Nucl. Phys. {\bf B202}, 327 (1982).
\item For a recent review and references to the original literature see
S. Raby, Prof. of the 1994 Trieste Summer School, hep-ph/9501349,
OSU Preprint OHSTPY-HEP-T-95-024 (1995).
\item F. Wilczek and A. Zee, Phys. Rev. {\bf D 25}, 553 (1982).
\item K.S. Babu and Q. Shafi, Phys. Rev. {\bf D 47}, 5004 (1993); Y.
Achiman and T. Greiner, Phys. Lett. {\bf B329}, 33 (1994); S. Naculich,
Phys. Rev. {\bf D 48}, 5293 (1993).
\item F. Gilman and Y. Nir, Annu. Rev. Part. Sci. {\bf 40}, 213 (1990);
K. Kang, J. Flanz and E. Paschos, Z. Phys. {\bf C 55}, 75 (1992); E. Ma,
Phys. Rev. {\bf D 43}, R2761 (1991).
\item A. Dannenberg, L. Hall and L. Randall, Nucl. Phys. {\bf B271}, 574
(1986); K.S. Babu and S.M. Barr, Phys. Rev. Lett. {\bf 72}, 2831 (1994).
\item  J. Harvey, P. Ramond and D. Reiss, Phys. Lett. {\bf B92}, 309
(1980); Nucl. Phys. {\bf B199}, 223 (1982);
S. Dimopoulos, L. Hall and S. Raby, Phys. Rev. Lett. {\bf 68}, 752
(1992); Phys. Rev. {\bf D 45}, 4192 (1992);
H. Arason, D. Castano,, E. Pirad and P. Ramond, Phys. Rev.
{\bf D 47}, 232 (1992);
V. Barger, M.S. Berger, T. Han and M. Zralek, Phys. Rev. Lett.
{\bf 68}, 3394 (1992).
\item K.S. Babu and R.N. Mohapatra, Bartol Preprint BA-94-56,
UMD-PP-95-57, hep-ph/9410326.
\item S.M. Barr, Phys. Rev. Lett. {\bf 64}, 353 (1990); Z. Berezhiani
and R. Rattazzi, Nucl. Phys. {\bf B 407}, 249 (1993).
\item G. Anderson, S. Dimopoulos, L. Hall, S. Raby and G. Starkman,
Phys. Rev. {\bf D 49}, 3660 (1994).
\item P. Ramond, R.G. Roberts and G.G. Ross, Nucl. Phys. {\bf B406}, 19
(1993); D. Kaplan and M. Schmaltz, Phys. Rev. {\bf D 49}, 3741 (1994);
C. Albright and S. Nandi, Phys. Rev. Lett. {\bf 73}, 930 (1994);
G. Leontaris and N. Tracas, Phys. Lett. {\bf B303}, 50 (1993).
\item K.S. Babu and Q. Shafi, Phys. Lett. {\bf B311}, 172 (1993);
S. Dimopoulos, L. Hall and S. Raby, Phys. Rev. {\bf D 47}, 3697
(1993).
\end{enumerate}

\section*{Figure Captions}
\begin{description}
\item {Fig. 1.}  A plot of $|V_{cb}|$ versus $|m_s/m_d|$ corresponding to
$\eta_P= +1$.  $\alpha_s(M_Z) = 0.12$ has been used along with
$m_c(m_c)=1.27~GeV$.  The three curves correspond to three different
values of $m_t^{\rm phys}$.

\item {Fig. 2.}  Same plot as in Fig. 1, but for $\eta_P = -1$.
\end{description}
\end{document}